\documentclass[prl,twocolumn,aps,english,showpacs,longbibliography,superscriptaddress]{revtex4-1}
\usepackage{natbib}

\usepackage{color}
\def\mathcolor#1#{\@mathcolor{#1}}
\def\@mathcolor#1#2#3{%
  \protect\leavevmode
  \begingroup\color#1{#2}#3\endgroup
}

\usepackage{graphicx}
\usepackage{dcolumn}
\usepackage{bm}
\usepackage{color}
\usepackage{tabularx}
\usepackage{amsmath,amsfonts,amsthm,amssymb}
\usepackage{appendix}

\usepackage{graphicx}
\usepackage{xcolor}
\definecolor{deeppink}{rgb}{1.0, 0.08, 0.58}

\usepackage{bm}
\usepackage{mathptmx}

\usepackage[english]{babel}

\usepackage[colorinlistoftodos]{todonotes}
\usepackage[colorlinks=true, allcolors=blue]{hyperref}

\begin{document}

\title{Manipulating the shape of flexible magnetoelastic nanodiscs with meron-like magnetic states}
% Force line breaks with \\
\author{Beatriz Miranda-Silva}
\affiliation{Departamento de F\'isica, Universidade Federal de Vi\c cosa, \\Avenida Peter Henry Rolfs s/n, 36570-000, Vi\c cosa, MG, Brasil}

\author{Pedro H. C. Taveira} 
%\email[Corresponding author: ]{rosa.corona@usach.cl}
\affiliation{Departamento de F\'isica, Universidade Federal de Vi\c cosa, \\Avenida Peter Henry Rolfs s/n, 36570-000, Vi\c cosa, MG, Brasil}

\author{Allison W. Teixeira}
\affiliation{Departamento de F\'isica, Universidade Federal de Vi\c cosa, \\Avenida Peter Henry Rolfs s/n, 36570-000, Vi\c cosa, MG, Brasil}

\author{Jakson M. Fonseca}
\affiliation{Departamento de F\'isica, Universidade Federal de Vi\c cosa, \\Avenida Peter Henry Rolfs s/n, 36570-000, Vi\c cosa, MG, Brasil}

%\author{Leandro G. Rizzi}
%\affiliation{\Ufv}

\author{Ricardo G. El\'ias}
% \homepage{http://www.Second.institution.edu/~Charlie.Author.}
\affiliation{Universidad de Santiago de Chile (USACH) Departamento de F\'isica, CEDENNA, Avda. Ecuador 3493, Estaci\'on Central, Santiago, Chile}%
\affiliation{Center for the Development of Nanoscience and Nanotechnology, Avda. Libertador Bernardo O'Higgins 3363, 9170124 Santiago, Chile}%

\author{Nicol\'as Vidal-Silva}
% \homepage{http://www.Second.institution.edu/~Charlie.Author.}
\affiliation{Departamento de Ciencias F\'isicas, Universidad de La Frontera, Casilla 54-D, 4811186 Temuco, Chile}%
%\affiliation{\cedenna}%

\author{Vagson L. Carvalho-Santos}
\affiliation{Departamento de F\'isica, Universidade Federal de Vi\c cosa, \\Avenida Peter Henry Rolfs s/n, 36570-000, Vi\c cosa, MG, Brasil}

\date{\today}% It is always \today, today,

\begin{abstract}
The control of the magnetic properties of shapeable devices and the manipulation of flexible structures by external magnetic fields is a keystone of future magnetoelectronics-based devices. This work studies the elastic properties of a magnetoelastic nanodisc that hosts a meron as the magnetic state and can be deformed from structures with positive to negative Gaussian curvature. We show that the winding number of the hosted meron is crucial to determine the curvature sign of the stable obtained shape. Additionally, we show that the optimum curvature reached by the nanodisc depends on geometrical and mechanical parameters. It is shown that an increase in the external radius, thickness, and Young's modulus lead to a decrease in the optimum curvature absolute value. Finally, it is shown that the nanodisc's shape also depends on the connection between the polarity and chirality of the vortex-like meron.    
\end{abstract}
\maketitle

\section{Introduction}

%Flexible systems are a key stone to compose future technological devices based on the concepts of shapeable magnetoelectronics. Indeed, due to 
The possibility to reshape electronic systems has promoted the concept of flexible and stretchable electronics to a hot topic in soft matter researches \cite{Makarov-PRAppl,Bujak-Chem} considering applications of shapeable systems in sensory devices \cite{Melzer-Nat,Vandeparre,Webb-Nat,Graudejus}, solar cells \cite{Lipomi}, electronic skins \cite{Melzer-Nat,Lumelsky,Hammok,Gu-Kim}, soft robotics \cite{WHu-Nature}, wearable devices \cite{Gomg-Nat}, and for manipulating the shape of liquid interfaces \cite{Timounay}. Also, the inclusion of the magnetic freedom degree into soft systems is fascinating because there are several possibilities to manipulate the magnetic properties by changing its shape or produce effects on the geometry of the system by applying external magnetic stimuli \cite{Makarov-PRAppl}. For instance, one can cite the possibility of using stretchable magnetoelectronics for the emerging field of soft robotics \cite{WHu-Nature}, and manipulating the shape of elastomeric actuators \cite{Stephan-ADM,Lum-PNAS}.

In general, most field-controllable materials with magnetically switchable properties consists of elastomers, that are magnetic nanoparticles embedded into a non-magnetizable polymer matrix \cite{Geryak}. Because the intrinsic properties of the magnetic particles do not affect the other ones, the proper description of the magnetic properties of elastomers is performed by considering a dipole-dipole interaction \cite{Romeis-PRE}. Therefore, when an external magnetic field is applied into the system, there is a particle rearrangement, which changes the mechanical properties of the elastomer matrix \cite{Coquele}. As a result, both the external field and initial arrangement of the magnetic particles influence the final stabilized shape of the magnetic elastomer \cite{Romeis-JMMM}. In this case, the competition between magnetic interactions with membrane bending and stretching can drive the membrane to expand, contract, or twist in such a way that many shapes can be obtained as a function of an external magnetic field \cite{Lum-PNAS,Brisbois-PNAS}.

Nevertheless, the imbibition of magnetic microparticles in an elastic matrix avoids the possibility of leading with magnetoelastic systems in the nanoscale range of sizes. This problem can be circumvented by constructing systems where a short-range exchange plays the role. Examples of such systems include organic, organic-inorganic hybrid, and molecule-based magnets, which exhibit different types of magnetic ordering \cite{Podgajny,Barrom-Nat,Miller-Chem,Miller-Tod}, even in a room-temperature environment \cite{Mahmood-Chem}. Therefore, due to the short-range of magnetic interactions, one can reduce the size of the flexible magnetic system whose shape can be manipulated by external magnetic fields. Theoretical works have considered magnetic subsystems where the short-range exchange interaction determines the magnetic properties of the particle, and stretching and bending are responsible for describing the energetic cost to deform the elastic subsystem. In this case, the nucleation of periodic solitons in the magnetic system induces the appearance of periodic shrinking of the membrane \cite{Dandoloff-1}, and curvature-induced geometrical frustration in magnetic systems in both cases under the absence and presence of external magnetic fields \cite{Dandoloff-2,Dandoloff-3,Vagson-BJP,Vagson-PLA}. Additionally, Yershov \textit{et al.}\cite{Yershov-Ring} showed that a unidimensional magnetoelastic ring presents a shape depending on the magnetic configuration. An onion or a vortex magnetic state leads the nanoring to assume an elliptical or circular shape, respectively. 
Finally, due to the intrinsic magnetochirality induced by the intrinsic Dzyaloshinskii-Moriya interaction (DMI), a flexible ribbon can be spontaneously deformed  \cite{Yershov-PRB-2019}. This ribbon deformation depends on the symmetry of the DMI and the mechanical, magnetic, and geometric parameters of the magnetoelastic body. 

Regarding magnetic properties, it is known that the introduction of curvature in quasi-2D systems induces effective interactions \cite{Gaididei-PRL} that are responsible, for instance, for a curvature-driven vortex \cite{Sloika-APL} and skyrmion \cite{Carvalho-2020,Kravchuk-2018} polarity-chirality connection, and a curvature-induced selection on the domain wall phase \cite{Yershov-PRB-2015,Cacilhas,Bittencourt}. The exchange-driven curvature induced polarity-chirality connection was evidenced by El\'ias \textit{et al.} \cite{Elias-SciRep}, that showed the existence of a curvature-induced winding number of merons hosted in rigid curved magnetic elements, where vortices and antivortices are nucleated in structures with positive and negative Gaussian curvatures, respectively. The authors also showed that the minimum energy configuration or the meron depends on the relative directions defined by the meron's polarity and chirality. Nevertheless, in that work, the range of considered parameters leads to minimum magnetic energy for the maximum (vortex) and minimum (antivortex) curvatures. If we consider that the considered structures are flexible, the new freedom degree brought by the elastic subsystem should yield an optimum value for the curvature minimizing the energy. Therefore, in this work, we propose the study of the static magnetic and mechanical properties of a magnetoelastic disc hosting a meron as the magnetic state. It is shown that the modulus of the optimum curvature of the stable structure decreases as the disc radius and thickness increase. As expected, the mechanical properties of the elastic subsystem also influence the obtained stable shape. Indeed, due to the increase in the structure rigidity, the optimum curvature decreases as Young's modulus increases. We also show that the proper control of the meron's chirality can be used to deform the shape of the disc from a paraboloidal structure, with positive curvature, to a saddle-like shape, with negative Gaussian curvature.

This work is divided as follow: Section \ref{TM} presents the adopted theoretical model to describe the magnetoelastic disc. In section \ref{RD} we present the obtained results and discussions. Section \ref{Conc} brings our conclusions and prospects. 

\section{Theoretical Model}\label{TM}

\begin{figure}
    \centering
    \includegraphics[width=8.5cm]{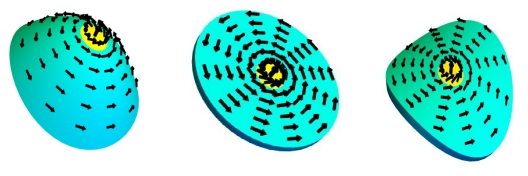}
    \caption{Schematic representation of the considered geometry with $R=1$ and  $h=0.1$ (a.u.). From left to right, we present the geometries for $c=0.8,\, 0, \, {\rm and} \, -0.5$ respectively. The vector field consists of a meron with $q=1$ and $\gamma=\pi/2$. Yellow region depicts a circumference of radius $r_0$, representing the meron's core.}
    \label{fig1}
\end{figure}

In this work, we analyze a magnetoelastic nanodisc, defined as a nanostructure with both magnetic and elastic degrees of freedom interacting, and exhibiting a meron-like configuration as a magnetic ground state. We assume that the nanodisc consists of a thin shell in such a way that its thickness $h$ is much smaller than the external radius $R$ of the disc ($h\ll R$). We also consider that the nanodisc can deform from structures having positive (paraboloidal surface) to a negative (saddle surface) Gaussian curvature. The geometrical description of the considered system can be given by

\begin{eqnarray}\label{Param}
%\mathbf{r}=x\mathbf{\mathbf{x}}+y\mathbf{\mathbf{y}}+c(cx^2+y^2)\mathbf{\mathbf{z}}\,,
\mathbf{r}=x\mathbf{\hat{x}}+y\mathbf{\hat{y}}+c(cx^2+y^2)\mathbf{\hat{z}}\,,
\end{eqnarray}

\noindent
where $x=\rho\,\cos\phi$, $y=\rho\,\sin\phi$, $\rho\in[0,R]$, $\phi\in[0,2\pi]$, $\{\mathbf{x},\mathbf{y},\mathbf{z}\}$ corresponds to the unitary vectors of the three-dimensional Cartesian space, and $c\in[-1,1]$ is a number that determines the surface curvature. That is, $c<0$ describes a hyperbolic paraboloid, that has a negative Gaussian curvature, and $c>0$ defines a paraboloidal surface, presenting a positive Gaussian curvature. If $c=0$, the parametrization describes a planar nanodisc. Fig. \ref{fig1} depicts the shapes of the magnetoelastic structures for the cases $c=0.8$, $c=0$, and $c=-0.5$.  

To properly describe the magnetic properties of the nanodisc, we use the micromagnetic approach, in which the magnetization is a continuous function of the position inside the magnetic element. Therefore, the magnetization field can be parametrized as a spherical coordinate system lying in a curvilinear basis, that is,

\begin{equation}\label{MagPar}
   % \mathbf{M} = \cos \Phi \sin \Theta \boldsymbol{\vec{\rho}} +
%    \sin \Phi \sin \Theta \boldsymbol{\vec{\phi}} +
 %   \cos \Theta \mathbf{n}\,,
  \mathbf{M} = \cos \Phi \sin \Theta \boldsymbol{\vec{\rho}} +
    \sin \Phi \sin \Theta \boldsymbol{\vec{\phi}} +
    \cos \Theta \boldsymbol{\vec{n}}\,,
\end{equation}
\noindent 
where $\boldsymbol{\vec{\rho}}=\mathbf{g}_\rho/\left\|\mathbf{g}_\rho\right\|$ and $\boldsymbol{\vec{\phi}}=\mathbf{g}_\phi/\left \|\mathbf{g}_\phi\right\|$ are unitary vectors pointing along the tangential direction on the surface of the structure, and $\boldsymbol{\vec{n}}=\boldsymbol{\vec{\rho}}\times\boldsymbol{\vec{\phi}}$ is the unitary vector pointing perpendicularly to the surface. Because the parametrization given in Eq. \eqref{Param} yields a non-orthogonal basis, the radial direction in the surface $\boldsymbol{\vec{\rho}}$ is not necessarily orthogonal to $\boldsymbol{\vec{\phi}}$. Here, we define the natural tangential basis $\mathbf{g}_\mu=\partial_\mu\mathbf{r}$ with $\partial_\mu\equiv \frac{\partial}{\partial x_i},\, i=1,\,2$, from which one can obtain the metric tensor elements $g_{\mu\nu}=\mathbf{g}_\mu\cdot\mathbf{g}_\nu$. 

In the considered parametrization, $\mathbf{M}$ has not in general a constant modulus, and then, it is convenient to define the normalized magnetization, given by
$\mathbf{m} = {\mathbf{M}}/{\left \| \mathbf{M} \right \|}$, with $\left \|\mathbf{M} \right\| =\sqrt{1+\sin \left(2\Phi\right)\left(\cos^2\Theta\right)\boldsymbol{\vec{\rho}}\cdot\boldsymbol{\vec{\phi}}}$. Following the ideas presented in Ref. [\onlinecite{Elias-SciRep}], we assume that the magnetization pattern of the nanodisc consists of a meron-like configuration, which can be well described by the following ansatz \cite{landeros2005}

\begin{equation}
\begin{split}
       \Theta (\rho) &= \arccos \left( \frac{p}{1+(\frac{\rho}{r_0})^s} \right) \\
    \Phi(\phi)  &= (q-1)\phi + \gamma
\end{split}
\end{equation}

\noindent
where $q$ is the winding number of the magnetic configuration, and represents the curl of the field around the meron’s core when projecting the field onto the surface. Therefore, a vortex or anti-vortex structure can be described  for $q = 1$ or $q=-1$, respectively. The meron is also characterized by the polarity $p$ of the core, which can be $1$ or $-1$ when the central magnetic moment points parallel or antiparallel to $\mathbf{n}$. The meron's topological charge is defined using the polarity and the winding number as \cite{Elias-PRB} $Q = pq/2$. The chirality of the meron is determined by the parameter $\gamma$, which consists of a phase that gives the orientation of the field with respect to the radial direction $\boldsymbol{\vec{\rho}}$ on the surface. The parameter $s$ is a positive integer that determines the meron's core size, with radius $r_0$, and is defined as the minimum radial distance between the surface's center and that when $\mathbf{\vec{m}}\cdot\boldsymbol{\vec{n}}=0$ occurs. The described magnetization vector field hosted in structures with different curvatures is presented in Fig. \ref{fig1} for a meron with $q=1$, $\gamma=\pi/2$, and $p=1$. The meron's core is represented by the smaller yellow circle. It is worth noticing that the adopted ansatz describing the meron's profile defines a vortex (antivortex) as a configuration that lies asymptotically in the in-surface plane.   

The total energy of the system is given by the sum between magnetic ($E_m$) and elastic ($E_{el}$) contributions, that is, $E=E_m+E_{el}$. Since we are dealing with very thin shells, we can approximate the dipolar energy by an easy-surface anisotropy. Therefore, the magnetic contribution to the total energy is determined by the exchange and anisotropy, given respectively by

\begin{equation}
    E_{x} = Ah \int g^{\mu \nu} \frac{\partial \mathbf m}{\partial x_\mu} 
    \cdot \frac{\partial \mathbf{m}}{\partial x_\nu} \sqrt{g}\,dx_\mu\,dx_\nu
\end{equation}

\noindent
and

\begin{equation}
    E_{ani} = K_a h\int (\mathbf m \cdot \mathbf n)^2 \sqrt{g}\,d\rho\,d\phi\,,
\end{equation}

\noindent 
where the Einstein summation convention over repeated indices is assumed. The parameter $A$ is the exchange stiffness and $K_a>0$ is the easy-surface anisotropy constant. Additionally, $\sqrt{g}=\sqrt{\left \|\text{det}\,g_{\mu\nu}\right\|}$, and the covariant ($g_{\mu\nu}$) and contravariant ($g^{\mu\nu}$) metric elements are obtained from the parametrization given in Eq. \eqref{Param} (See Appendix \ref{Apendice} for details on the calculation of these geometrical quantities). The competition between exchange and anisotropy interactions characterizes a magnetic length $\ell= \sqrt{A/K_a}$, which determines the length scale of the system (and the core's size of the magnetic structures). 

The elastic energy is determined from the sum between the stretching and bending energies, given by \cite{Efrati}

\begin{eqnarray}
    E_{el}=\int \left[h\,w_s+h^3\,w_b\right]\sqrt{\overline{g}}\,d\rho\,d\phi\,,
\end{eqnarray}

\noindent
where the stretching and bending energy densities are respectively 

\begin{eqnarray}\label{Stretching}
    w_{s} = 
    \frac{Y}{8(1+\varsigma)}
    \left(
    \frac{\varsigma}{1-\varsigma} \overline{g}^{\alpha\beta}\overline{g}^{\gamma\delta}+
    \overline{g}^{\alpha\gamma}\overline{g}^{\beta\delta}
    \right)\nonumber\\
    \times
    (g_{\alpha\beta} - \overline{g}_{\alpha\beta})
    (g_{\gamma\delta} - \overline{g}_{\gamma\delta})
\end{eqnarray}

\noindent
and 

\begin{equation}\label{Bending}
    w_{b} = \frac{Y\,b_{\alpha\beta}b_{\gamma\delta}}{24(1+\varsigma)}\left(
    \frac{\varsigma}{1-\varsigma}\overline{g}^{\alpha\beta} \overline{g}^{\gamma\delta}
    + \overline{g}^{\alpha\gamma} \overline{g}^{\beta\delta}
    \right)\,.
\end{equation}

\noindent
Here, the parameters $Y$ and $\varsigma$ are Young’s modulus and Poisson ratio \cite{Masao}, respectively, and $b_{\mu\nu}=\mathbf{n}\cdot\partial_\mu\mathbf{g}_\nu$. In addition, $\overline{g}_{\mu\nu}$ is the metric tensor for a nanodisc without elastic tensions. Here, we  consider the planar nanostructure, with $\overline{g}_{\mu\nu}=\delta_{\mu\nu}$, as a reference metric. The explicit elastic energy densities expressions written in terms of the geometrical parameters of the considered structures are cumbersome and therefore, they are presented in Appendix \ref{Apendice}. An important point is that both magnetic and elastic subsystems interact through the anisotropy energy as this energy involves the normal direction of the magnetization field, but such a direction depends on the surface's geometry, which, in our model, can be deformed.

\section{Results}\label{RD}

\subsection{Winding-number-induced curvature}

The model described above allows us to determine the magnetoelastic properties of the considered structures by calculating the total energy as a function of different geometrical and material parameters. Following the ideas presented in Ref. [\onlinecite{Yershov-PRB-2019}], we do not consider a specific material but analyze the influence of a range of parameters in the possibility of controlling the shape of a magnetoelastic membrane through modifying its magnetization configuration. In all examined cases, the presented results regard the meron's core size that minimizes the magnetic energy. Additionally, the results presented in this section regards to merons with $p=+1$ and $\gamma=0$.

\begin{figure}
    \centering
    \includegraphics[width=8.5cm]{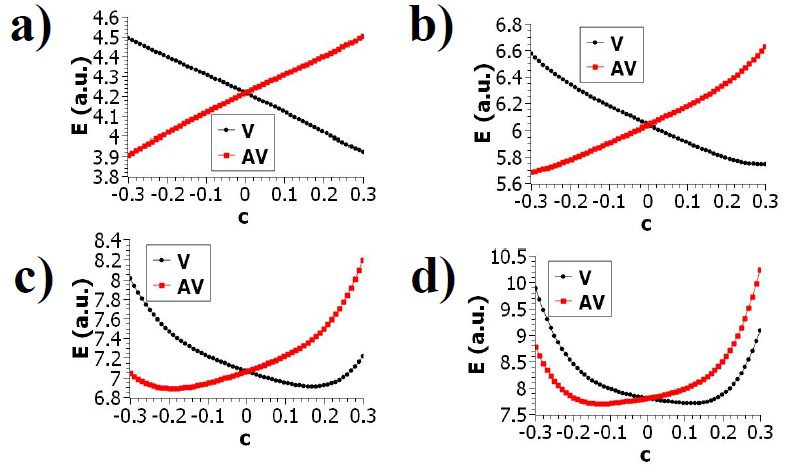}
    \caption{Total energy as a function of $c$ for different disc radius. Figures a, b, c, and d present the results for $R=4\ell,\,6\ell,\,8\ell$ and $10 \ell$, respectively. The thickness is $h = 0.05\ell$, and $A/(Yh^2)=1$.}
    \label{fig2}
\end{figure}

Firstly, we determine the behavior of the total energy as a function of the disc radius $R$ for a fixed value of $h$, and $A/(Yh^2)=1$. The obtained results are depicted in Fig. \ref{fig2} for $R=4\ell$, $6\ell$, $8\ell$, and $10\ell$, with the nanodisc height given by $h=0.05\ell$. In all cases, we observe that the meron's winding number determines the curvature's sign that minimizes the total energy, as previously predicted by El\'ias \textit{et al.}, that considered merons hosted in rigid structures, showing the emergence of a curvature-induced selection of the meron's winding number \cite{Elias-SciRep}. Nevertheless, in that case, the curvature that minimizes the total energy was independent of the external radius of the nanodisc. In the present study, the curvature that minimizes the total energy is radius-dependent because the elastic energy brings a new freedom degree to the system. From the analysis of Figs. \ref{fig2}-a and \ref{fig2}-b, the magnetoelastic structure deforms until its curvature reaches the maximum value in the considered range ($c=0.3$). This behavior can be explained by the high energy cost to nucleate the meron's core, which, for small radii, occupies a substantial area in the nanodisc in comparison with its external radius. Therefore, the structure presents a large deformation to diminish the anisotropic energy. As the radius of the nanostructure increases, the meron's core energy gives a lower contribution compared to other energy terms. Therefore, there is a limit in the maximum curvature that the magnetoelastic disc can reach. For instance, Fig. \ref{fig2}-d reveals that a nanodisc with an external radius of $10\ell$ can deform until it reaches a curvature of $c\approx0.14$ or $c\approx-0.17$ when it hosts a vortex or an antivortex, respectively.  

\begin{figure}
    \centering
    \includegraphics[scale=0.5]{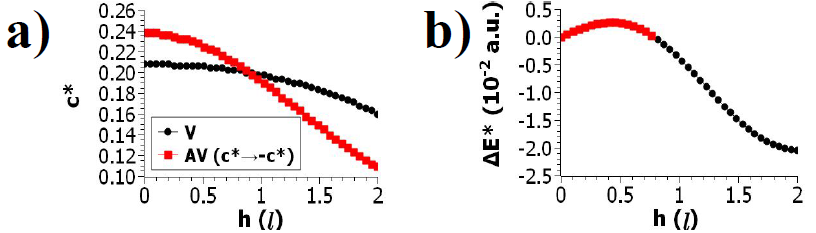}
    % deltaE* = V - AV
    \caption{Fig. a) depicts the optimum curvature $c^*$ as a function of the thickness $h$. Here the curvature for a system hosting an antivortex is given in modulus ($-c^*>0$). Fig. b) depicts the difference between the energies of a vortex and an antivortex (regarding their respective optimum curvatures) as a function of $h$. In both cases $h$ ranges from $h= 10^{-17} \ell$ until $h = 2\ell$.}  
    \label{fig3}
\end{figure}

Let us now explore the impact of changes on the geometry and mechanical properties in the deformation of the magnetoelastic membrane. Specifically, we are interested in studying the optimum curvature ($c^*$) dependence on the thickness of the nanostructure and the effect of changing Young's modulus, which is the parameter controlling the elastic stiffness of the material under external forces. Fig. \ref{fig3} shows the value of $c^*$ as a function of $h$ for a nanodisc hosting a vortex (Black squares) or antivortex (Red circles) with radius $R=8\ell$, and $A/(Yh)^2=1$. To provide a graphic richer in details, in Fig. \ref{fig3}-a, we present the modulus of the optimum curvature for a system hosting an antivortex and the optimum curvature for a vortex. That is, the presented values of $c^*$ for the antivortex configuration should be read as $c^*\rightarrow-c^*$. Again, the vortex and antivortex yield a deformation in the nanodisc leading to a positive and negative curvature, respectively. It can be noticed that the increase in the disc thickness leads to a reduction in the absolute value of the curvature for which the vortex and antivortex patterns reach the minimum energy value. The reduction in the nanostructure curvature as $h$ increases is associated with the higher elastic cost to deform thicker structures (note that $E_{el}$ scales as $h^3$). We have also determined the total energy $E(c = c^*)\equiv E^*$ (using the obtained values of $c^*$) as a function of $h$. Fig. \ref{fig3}-b represents the obtained results for the energy difference $(E_{V}^*-E_{AV}^*)$ between a system hosting an antivortex and a vortex magnetization profile. It can be observed that there is a thickness-induced selection of meron winding numbers in the system. That is, for $h\lesssim 0.8\ell$ (region represented by red squares in Fig. \ref{fig3}-b),  an antivortex hosted in a structure with negative curvature presents lower energy than a vortex in a nanostructure with positive curvature. Nevertheless, for $h\gtrsim0.8\ell$ (region represented by black dots in Fig. \ref{fig3}-b), the lower energy state consists of a vortex hosted in a parabolic-shaped system.

\begin{figure}[h]
    \centering
    \includegraphics[width=8.5cm]{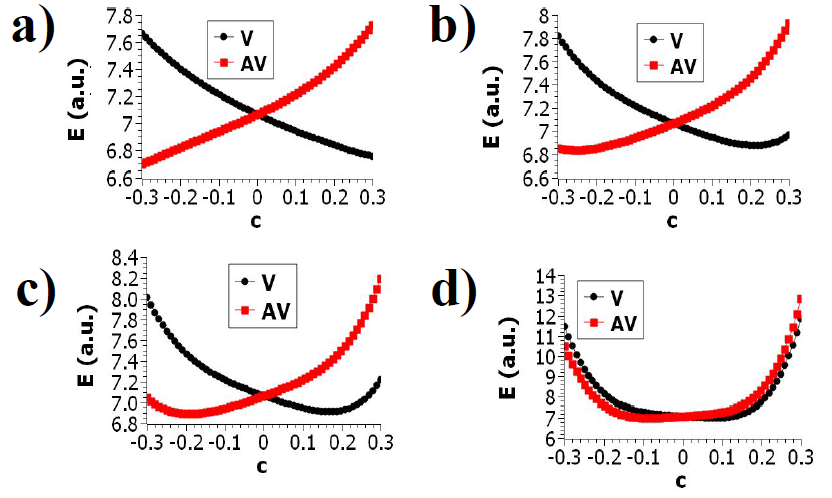}
    % Para raio R=0.8 e h=0.005
    % a) Y = 40
    % b) Y = 200
    % c) Y = 400
    % d) Y = 4000
    \caption{System energy as a function of the curvature for a nanodisc with radius $R=8\ell$ and $h=0.05\ell$ for different Young's modulus. Figures a, b, c, and d, depict a nanostructure with $A/(Yh^2)=10,\, 5,\, 1$ and $0.1$ respectively.}
    \label{fig4:Young}
\end{figure} 

We also analyze the effects of Young's modulus on the shape of the nanostructure. In this context, in Fig. \ref{fig4:Young} we depict the total energy of the system as a function of the curvature for distinct values of $Y$. As expected, for small Young's modulus ($A/(Yh^2)=10$), the reached optimum curvature does not appear in the range of parameters considered here, evidencing that the nanostructure deforms until it reaches high values of $|c|$ (See Fig. \ref{fig4:Young}-a). Figs. \ref{fig4:Young}-b and \ref{fig4:Young}-c present the obtained results for $A/(Yh^2)=5$ and $1$, respectively. One can notice that the optimum curvature of the system decreases as the Young's modulus increases. This behavior is similar to that one observed when we have varied the thickness of the nanodisc. That is, the curvature that minimizes the energy diminishes in modulus, indicating the increase in the structure rigidity. For $A/(Yh^2)\lesssim0.1$ (Fig. \ref{fig4:Young}-d), the elastic energy cost to deform the nanostructure is very high, and the nanodisc shape becomes independent of the winding number of the hosted meron, which eventually reads the optimum curvature $\vert c^*\vert\approx0$.

\begin{figure}[h]
    \centering
    \includegraphics[width=8.5cm]{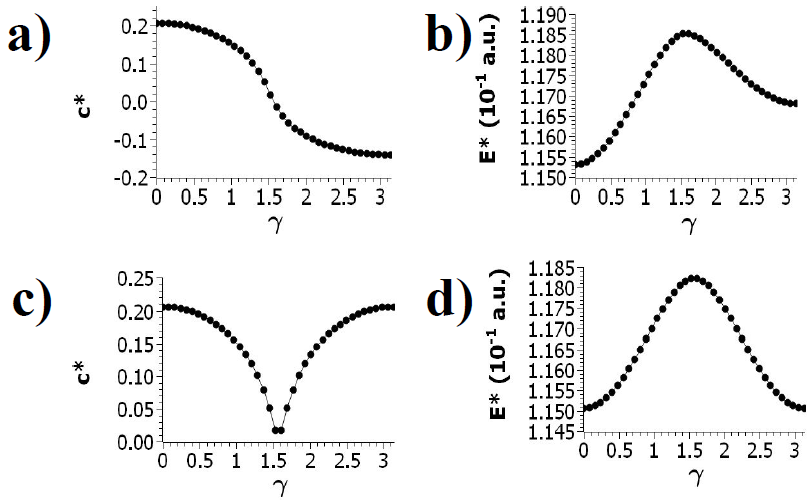}
    % Graph of optimum curvature as a function of the quirality.
    \caption{optimum curvature (a and c) and total energy (b and d) as a function of $\gamma$ for a unchanged (top graphics) and variable (bottom graphics) polarity.}
    \label{fig5}
\end{figure} 

\subsection{Influence of chirality on curvature}

A remarkable feature of curved magnetic shells is the appearance of exchange-driven effective anisotropy and Dzyaloshinskii-Moriya interactions \cite{Gaididei-PRL}, which are responsible, for instance, for the increase in the skyrmion stability in hills and valleys \cite{Carvalho-2020,Kravchuk-2018}, and the emergence of curvature-induced forces in particle-like magnetization configurations \cite{Carvalho-APL-2021,Yershov-PRB-2018}. Additionally, such effective interactions yield a curvature-induced phase selection in the domain wall (DW) phase, where the kind of DW head-to-head or tail-to-tail are always directed outward and inward the bend, respectively \cite{Yershov-PRB-2015}. Moreover, this phase selectivity also determines a polarity-chirality connection, where depending on the vortex chirality, there is a preferential direction for where its core points \cite{Elias-SciRep,Sloika-APL}. Based on the above, we explore the possibility of manipulating the shape of the magnetoelastic particle by using meron-like configurations with a fixed polarity $p=1$ (outward) and controlling the vortex chirality. In Fig. \ref{fig5} we have obtained both the optimum curvature $c^*$ and the total energy $E^*$ as a function of $\gamma$ for a nanodisc with $R=8\ell$, $h=0.05\ell$, and $A/(Yh)^2=1$. The analysis of Fig. \ref{fig5}-a reveals that by changing the vortex chirality there is a change in the shape of the nanostructure, where we can notice a change in the curvature sign when $\gamma=\pi/2$. Therefore, the magnetoelastic system changes its shape from a paraboloid-like to a saddle-like geometry. To understand this chirality-induced shape change, we have determined the total energy as a function of $\gamma$, whose results are presented in Fig. \ref{fig5}-b, revealing that the energy increases with $\gamma$. Thus, hosting a vortex state for certain values of $\gamma$, the system is deformed in a negatively curved surface to control the increase in the magnetic energy due to the polarity-chirality connection \cite{Elias-SciRep,Sloika-APL}. To prove the above statement, we have analyzed the optimum curvature and the total energy as a function of $\gamma$, but now, we consider a change in the vortex polarity from $p=1$ to $p=-1$ when $\gamma=\pi/2$. In this case, as presented in Figs. \ref{fig5}-c and \ref{fig5}-d, we observe that the vortex polarity change makes the system remains with positive curvature.

\begin{figure}[h]
    \centering
    \includegraphics[width=8.5cm]{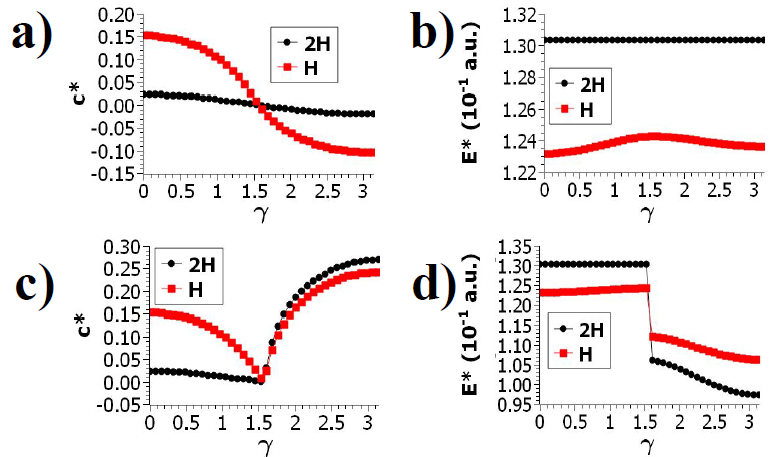}
    % fig6_2 : with energy
    \caption{optimum curvature and total energy for fixed (a and b) e variable (c and d) polarity as a function of the vortex chirality for a nanodisc under the action of a magnetic field $\mathbf{B}=H\mathbf{\hat{z}}$.}
    \label{fig6}
\end{figure}

An external magnetic field can drive the proper control of the vortex chirality. Therefore, we have analyzed the effects that a uniform magnetic field applied along the $z$-axis direction produces on the shape of the nanodisc hosting a vortex as a magnetization state. In this context, we determine the optimum curvature as a function of the vortex chirality for a magnetoelastic nanodisc under the action of an external magnetic field $\mathbf{B}=-H\mathbf{\hat{z}}$. Fig. \ref{fig6}-a and \ref{fig6}-b show, respectively, the optimum curvature and total energy as a function of the vortex chirality ($\gamma$) when we fix its polarity $p=+1$. Red squares and black dots show the behavior of a nanodisc under the action of a magnetic field with $\sqrt{H/K_a}=0.1$ and $0.14$, respectively. It can be noticed that the obtained optimum curvature is a result of the competition between the polarity-chirality connection induced by curvature effects and Zeeman interaction. That is, a direct comparison between Fig. \ref{fig5}-a and \ref{fig6}-a evidences that the magnetic field forces the system to diminish the modulus of its optimum curvature far all values of $\gamma$. Nevertheless, because the magnetic field direction favors the magnetic moments pointing along $-\mathbf{z}$ direction, it can be observed that the polarity changes when $\gamma\geq\pi/2$, being the optimum curvature positive for any value of $\gamma$, and increases as  function of the magnetic field (See Fig. \ref{fig6}-c). Finally, as expected, when the vortex polarity changes and points along the magnetic field direction, the total energy decreases, evidencing that this state minimizes both Zeeman and exchange interactions in the magnetic subsystem (See Fig. \ref{fig6}-d). \\

\section{Conclusions}\label{Conc}

We studied the properties of a magnetoelastic nanodisc hosting merons as a magnetic state for a large range of geometrical, elastic, and magnetic parameters. It is obtained that in the absence of an external magnetic field, the meron's winding number curvature determines if the nanoparticle's shape presents a positive or negative curvature. Additionally, the optimum curvature adopted by the nanodisc is a function of its radius, thickness, and Young's modulus. It was obtained that the absolute value of the optimum curvature decreases with both Young's modulus and disc thickness. We also showed that due to the exchange-driven curvature-induced effective DMI, changes in the vortex chirality yield changes in the nanoparticle's shape. Finally, it was shown that external magnetic fields can be used to change the nanoparticle's shape by the proper control of the vortex chirality. Indeed, because uniform magnetic fields favor the parallel magnetic moments alignment, the lower energy state lower energy is obtained when the Zeeman energy is favored while respecting the polarity-chirality connection.\\

It is worth noticing that the above-described results does not describe the shape dynamical evolution of the magnetoelastic disc under the action of a magnetic field, but present several static properties of magnetoelastic discs hosting merons as magnetization states.

\section*{Acknowledgments}
In Brazil, we thank the financial support of FAPEMIG and CNPq (Grant numbers 302084/2019-3 and 309484/2018-9). N. V.-S. thanks Fondecyt Postdoctorado Grant No. 3190264.\\

\begin{widetext}
\appendix

\section{Geometrical aspects of the considered structures and expressions for energy densities}\label{Apendice}

\subsection{Metric Elements}
The considered structures are parametrized by Eq. \eqref{Param}, which allows us to obtain the covariant metric matrix, given by

\begin{equation}
    g = 
\begin{pmatrix}
g_{11} &  g_{12}\\ 
g_{21} & g_{22}
\end{pmatrix}\,,
\end{equation}

\noindent
where

\begin{equation}\label{Metric}
    \left\{\begin{matrix}
    \begin{split}
g_{11} &= 1+4\rho^2c^2(c\cos^2(\phi) + \sin^2\phi)^2 \\ 
g_{12} &= 2(1-c)\rho^3c^2(c\cos^2\phi + \sin^2\phi)) \\ 
g_{21} &= g_{12}\\ 
g_{22} &= \rho^2[1+(1-c)^2\rho^2c^2\sin^2(2\phi) ]
\end{split}
\end{matrix}\right.
\end{equation}

It can be noticed that the metric elements of the reference system ($\overline{g}$) is obtained from taking $c=0$ (planar disc) in above equation. Therefore, we obtain $\overline{g}_{\rho\rho}=1$, $\overline{g}_{\phi\phi}=\rho^2$, and $\overline{g}_{\rho\phi}=\overline{g}_{\phi\rho}=0$. The contravariant metric elements $g^{\mu\nu}$ can be obtained from the relation $g^{\mu\nu}g_{\mu\nu}=\delta^{\mu}_{\nu}$. 

\subsection{Elastic energy densities}

\subsubsection{Stretching energy}

The stretching energy consists of the cost to produce changes in the size of the elastic subsystem. In the adopted theoretical model, the stretching energy density can be obtained from substituting the metric elements given in Eq. \eqref{Metric} in Eq. \eqref{Stretching}. After some algebra, we obtain 

\begin{equation}
\begin{split}
    w_{s} = &\frac{Y}{8(1+\nu)}\bigg \{
    \frac{16\rho^4c^4(c\cos^2(\phi) + \sin^2\phi)}{1-\nu}+
    8\rho^2 [(1-c)\rho^3c^2(c\cos^2\phi + \sin^2\phi))]^2
    \\
    &+
    \frac{2\nu \rho^2}{1-\nu} 
    (4\rho^2c^2\sqrt{c\cos^2(\phi) + \sin^2\phi})(\rho^2(1+(1-c)^2\rho^2c^2\sin^2(2\phi) )- \rho^{-2})\\
    &+ 
    \frac{\rho^4}{1-\nu}[\rho^2(1+(1-c)^2\rho^2c^2\sin^2(2\phi) )-\rho^{-2}]^2
    \bigg \}\,.
\end{split}
\end{equation}

\subsubsection{Bending energy density}

The introduction of bents in the elastic subsystem also yields an energetic cost, so called bending energy, whose energy density is given in Eq. \eqref{Bending}, where the matrix elements $b_{\mu\nu}$ are evaluated as

\begin{equation}
    \left\{\begin{matrix}
    \begin{split}
b_{\rho\rho} &= \frac{2c(c\cos^2\phi+\sin^2\phi)}{[1 + {4c^4 \rho^2\cos^2\phi}
+ 4 c^2 \rho^2\sin^2\phi]^{1/2}}
\\ 
b_{\rho\phi} &=b_{\phi\rho}= \frac{(1-c)c\rho\sin{(2\phi)}}{[1+ 4c^4 \rho^2 \cos^2 \phi +4c^2\rho^2\sin^2\phi]^{1/2}}
\\ 
b_{\phi\phi} &= \frac{2\rho^2[c^2 \cos^2 \phi + c \sin^2 \phi + 2c (1-c)\cos(2\phi)]}{[1+ 4c^4 \rho^2 \cos^2 \phi +4c^2\rho^2\sin^2\phi]^{1/2}}\,.
\end{split}
\end{matrix}\right.
\end{equation}

\noindent
Therefore, the substitution of the metric elements and the matrix elements $b+{\mu\nu}$ in Eq. \eqref{Bending} leads to the bending energy density written as

\begin{equation}
\begin{split}
    w_{b}(\rho,\phi) = &\frac{Y}{24(1+\nu)[1+4c^2\rho^2(c^2\cos^2\phi+ \sin^2\phi)]}\bigg \{
 \left( \frac{4}{1-\nu} \right) [(\mathcal{C} + 2c(1-c)\cos(2\phi))^2 + \mathcal{C}^2]
    \\
    &+ \frac{2\nu}{1-\nu}[4\mathcal{C}(c^2\cos^2\phi + c\sin^2 \phi + 2 c 
    (1-c)\cos(2\phi)]
 + 2(1-c)^2 c^2 \sin^2(2\phi)
    \bigg \}\,,
\end{split}
\end{equation}

\noindent
where $\mathcal{C} = c^4 (c \cos^2 \phi + \sin^2 \phi)^2$.

\end{widetext}

\end{document}